\documentclass[12pt]{article}
\usepackage{amssymb}
\renewcommand{\theequation}{\arabic{section}.\arabic{equation}}

\begin{document}



\def\a{\alpha}
\def\b{\beta}
\def\d{\delta}
\def\e{\epsilon}
\def\g{\gamma}
\def\h{\mathfrak{h}}
\def\k{\kappa}
\def\l{\lambda}
\def\o{\omega}
\def\p{\wp}
\def\r{\rho}
\def\t{\theta}
\def\s{\sigma}
\def\z{\zeta}
\def\x{\xi}
 \def\A{{\cal{A}}}
 \def\B{{\cal{B}}}
 \def\C{{\cal{C}}}
 \def\D{{\cal{D}}}
\def\G{\Gamma}
\def\K{{\cal{K}}}
\def\O{\Omega}
\def\R{\bar{R}}
\def\T{{\cal{T}}}
\def\L{\Lambda}
\def\f{E_{\tau,\eta}(sl_2)}
\def\E{E_{\tau,\eta}(sl_n)}
\def\Zb{\mathbb{Z}}
\def\Cb{\mathbb{C}}

\def\R{\overline{R}}

\def\beq{\begin{equation}}
\def\eeq{\end{equation}}
\def\bea{\begin{eqnarray}}
\def\eea{\end{eqnarray}}
\def\ba{\begin{array}}
\def\ea{\end{array}}
\def\no{\nonumber}
\def\le{\langle}
\def\re{\rangle}
\def\lt{\left}
\def\rt{\right}

\newtheorem{Theorem}{Theorem}
\newtheorem{Definition}{Definition}
\newtheorem{Proposition}{Proposition}
\newtheorem{Lemma}{Lemma}
\newtheorem{Corollary}{Corollary}
\newcommand{\proof}[1]{{\bf Proof. }
        #1\begin{flushright}$\Box$\end{flushright}}

\baselineskip=20pt

\newfont{\elevenmib}{cmmib10 scaled\magstep1}
\newcommand{\preprint}{
   \begin{flushleft}
   \end{flushleft}\vspace{-1.3cm}
   \begin{flushright}\normalsize
   \end{flushright}}
\newcommand{\Title}[1]{{\baselineskip=26pt
   \begin{center} \Large \bf #1 \\ \ \\ \end{center}}}
\newcommand{\Author}{\begin{center}
   \large \bf
Wen-Li Yang ${}^{a,b}$, and
  Yao-Zhong Zhang ${}^{b}$ \end{center}}
\newcommand{\Address}{\begin{center}

${}^a$ Institute of Modern Physics, Northwest University,
       Xian 710069, P.R. China\\
${}^b$ The University of Queensland, School of Physical Sciences,
Brisbane,
       QLD 4072, Australia\\


   \end{center}}
\newcommand{\Accepted}[1]{\begin{center}
   {\large \sf #1}\\ \vspace{1mm}{\small \sf Accepted for Publication}
   \end{center}}

\preprint
\thispagestyle{empty}
\bigskip\bigskip\bigskip

\Title{Free-field realization of the $osp(2n|2n)$ current algebra}
\Author

\Address

\vspace{0.28truecm}

{\it\large
 \begin{center}
  Dedicated to Professor Ryu Sasaki for his sixtieth birthday
 \end{center}}

\vspace{0.8truecm}

\begin{abstract}
The $osp(2n|2n)$ current algebra for a {\it generic} positive
integer $n$ at general level $k$ is investigated. Its free-field
representation and corresponding energy-momentum tensor are
constructed. The associated screening currents of the first kind
are also presented.

\vspace{1truecm} \noindent {\it PACS:} 11.25.Hf

\end{abstract}
\newpage
\section{Introduction}
\label{intro} \setcounter{equation}{0}

The interest in conformal field theories (CFTs) \cite{Bel84,Fra97}
related to superalgebras has grown  over the last ten years
because of their applications in physics areas ranging from string
theory \cite{Ber99,Bers99} and logarithmic CFTs \cite{Roz92,Gur93}
(for a review, see e.g. \cite{Flo03,Gab03}, and references
therein) to modern condensed matter physics
\cite{Efe83,Ber95,Mud96,Maa97,Bas00,Gur00}. In particular,  the
Wess-Zumino-Novikov-Witten(WZNW) models associated with the
supergroups $GL(n|n)$ and $OSP(2n|2n)$ stand out as an important
class of CFTs due to the fact that they have vanishing central
charge and primary fields with negative dimensions
\cite{Ber95,Mud96,Lud00,Bha01,Sch06}. However, unlike with their
bosonic versions,  the WZNW models on supergroups are far from
being {\it understood} \cite{Sch06} (for references therein and
some recent progress see \cite{Que07}), largely due to technical
reasons such as indecomposability of the operator product
expansion (OPE), appearance of logarithms in correlation functions
and continuous modular transformations of the irreducible
characters \cite{Sem03}.

On the other hand, the Wakimoto free-field realizations of current
algebras (or affine algebras \cite{Kac90}) \cite{Wak86} have been
proven to be  powerful in the study of the WZNW models on bosonic
groups \cite{Dos84,Fat86,God86,Ber90,Fur93,And95}. The free-field
realizations of bosonic current algebras have been extensively
investigated
\cite{Fei90,Bou90,Ber86,Ber89,Ito90,Ger90,Fre94,Boe97,Yan08}.
However, to our knowledge explicit free-field expressions for
current algebras associated with superalgebras  have so far been
known only for some isolated cases: those associated with
superalgebrs $gl(2|1)$ and $gl(2|2)$
\cite{Bow96,Ras98,Din03,Din03-1}, $gl(m|n)$ \cite{Yan07},
$osp(1|2)$ and $osp(2|2)$ \cite{Ber89,Ras98}. In particular,
explicit free-field expressions of the $osp(2n|2n)$ current
algebra for generic $n$  are still lacking due to the fact that it
practically would be very {\it involved} (if not impossible) to
apply the general procedure developed in \cite{Ito90,Ras98} to the
current algebra with a large $n$.

The recent studies  of the random bond Ising model have revealed
 that an appropriate algebraic framework
for studying multispecies Gaussian disordered systems at
criticality is based on the $osp(2n|2n)$ current algebra with the
positive integer $n$ related to the species
\cite{Ber95,Maa97,Gur00}. The explicit free-field expressions of
the $osp(2n|2n)$ current algebra will enable one to explicitly
construct correlation functions \cite{Dos84,Ber90,Fra97} of the
disordered systems at the critical points.

In this paper, motivated by its great applications to both string
theory and condensed matter physics, we investigate the
$osp(2n|2n)$ current algebra associated with the $OSP(2n|2n)$ WZNW
model at general level $k$ for a {\it generic} positive integer
$n$. Based on a particular order introduced in \cite{Yan08-1} for
the roots of (super)algebras we work  out the explicit expression
of the differential realization of $osp(2n|2n)$. We then apply the
differential realization to construct explicit free-field
representation of the current algebra. This representation
provides the Verma modules of the algebra.

This paper is organized as follows. In section 2, we briefly
review the definitions of finite-dimensional superalgebra
$osp(2n|2n)$ and the associated current algebra, which also
introduces  our notation and some basic ingredients. In section 3,
after constructing explicitly the differential operator
realization of $osp(2n|2n)$,  we construct the explicit free-field
representation of the $osp(2n|2n)$ current algebra at a generic
level $k$. In section 4, we construct the free-field realization
of the corresponding energy-momentum tensor by the Sugawara
construction. In section 5, we construct the free-field
realization of the associated screening currents. Section 6 is for
conclusions. In the Appendix A, we give the matrix realizations
associated with the defining representation for all generators of
$osp(2n|2n)$.


\section{Notation and preliminaries}
\label{CUR} \setcounter{equation}{0}

Let us fix our notation for the underlying non-affine superalgebra
$osp(2n|2n)$ for a positive integer $n$.\, $osp(2n|2n)$ is a
$\Zb_2$-graded  simple  superalgebra with a dimension $d=8n^2$.
Let $\lt\{E_i|i=1,\ldots,d=8n^2\rt\}$ be a basis of $osp(2n|2n)$
with a definite $\Zb_2$-grading and denoting the corresponding
grading by $[E_i]$. The generators $\lt\{E_i\rt\}$ satisfy
(anti)commutation relations, \bea
 \lt[E_i,\,E_j\rt]=\sum_{m=1}^{8n^2}f_{ij}^m\,E_m,\label{construct-constant}
\eea where $f_{ij}^m$ are the structure constants of $osp(2n|2n)$.
Here and throughout, we adopt the convention:
$[a,b]=ab-(-1)^{[a][b]}ba$, which extends to inhomogenous elements
through linearity.

One can introduce  a nondegenerate and invariant supersymmetric
metric or bilinear form $\lt(E_i,E_j\rt)$ for $osp(2n|2n)$ by
(\ref{Bilinear-D}). Then the $osp(2n|2n)$ current algebra (or
affine algebra $osp(2n|2n)_k$ \cite{Kac90}) is generated by the
currents $E_i(z)$ associated with the generators $E_i$ of
$osp(2n|2n)$. The current algebra at a general level $k$ obeys the
following OPEs \cite{Fra97}, \bea
 E_i(z)E_j(w)=k\frac{(E_i,E_j)}{(z-w)^2}
 +\frac{\sum_{m=1}^{d}f_{ij}^mE_m(w)}{(z-w)},\qquad i,j=1,\ldots,d,
 \label{current-OPE}
\eea where $f_{ij}^m$ are the structure constants in
(\ref{construct-constant}) and $d$ is the dimension of
$osp(2n|2n)$, i.e. $d=8n^2$. The aim of this paper is to construct
explicit free-field realizations of the $osp(2n|2n)$ current
algebra with a generic positive integer $n$ at a generic level
$k$.

Alternatively, one can  use the associated root system to label
the generators of $osp(2n|2n)$ as follows \cite{Fra96}. Because of
the fact that the rank of $osp(2n|2n)$ is $2n$, let us introduce
$2n$ linear-independent vectors: $\{\d_i|i=1,\ldots,n\}$ and
$\{\e_i|i=1,\ldots n\}$. These vectors are endowed a symmetric
inter product such that \bea
 (\d_m,\d_l)=\d_{ml},\quad (\d_m,\e_i)=0,\quad
 (\e_i,\e_j)=-\d_{ij} ,\qquad
 i,j,m,l=1,\ldots,n.\label{Inter-product}
\eea The set of roots of $osp(2n|2n)$ (or $D(n,n)$), denoted by
$\Delta$, can be expressed in terms of
$\lt\{\d_i,\,\e_j|i,j=1,\ldots,n\rt\}$ as
\bea
 \Delta=\lt\{\pm\e_i\pm\e_j,\,\pm\d_m\pm\d_l,\,\pm2\d_l,\,\pm\d_l\pm\e_i\rt\},
 \qquad i\neq j,\,\,m\neq l,
\eea while the set of even roots denoted by $\Delta_{\bar{0}}$ and
the set of odd roots denoted by $\Delta_{\bar{1}}$ are given
respectively by \bea
 \Delta_{\bar{0}}=\lt\{\pm\e_i\pm\e_j,\,\pm\d_m\pm\d_l,\,\pm2\d_l\rt\},
 \quad \Delta_{\bar{1}}=\lt\{\pm\d_l\pm\e_i\rt\},\qquad i\neq j,\,\,m\neq
 l.
\eea The distinguished simple roots are
 \bea
 &&\a_1=\d_1-\d_2,\ldots,
    \a_{n-1}=\d_{n-1}-\d_n,\,\a_n=\d_{n}-\e_1,\no\\
 &&\a_{n+1}=\e_1-\e_2,\dots,\a_{2n-1}=\e_{n-1}-\e_n,\,
   \a_{2n}=\e_{n-1}+\e_{n}.\label{simple-roots-D(r,n)}
\eea Then the corresponding positive roots denoted by $\Delta_+$
are
\bea
 &&\d_m-\d_l,\quad 2\d_l,\quad \d_m+\d_l,\quad 1\leq m<l\leq n,\label{positive-roots-1}\\
 &&\d_l-\e_i,\quad \d_l+\e_i,\qquad 1\leq i,l\leq n,\\
 &&\e_i-\e_j,\quad \e_i+\e_j,\qquad 1\leq i<j\leq n.\label{positive-roots-2}
\eea Among these positive roots, $\{\d_l\pm\e_i|i,l=1\ldots,n\}$
are odd and the others are even. Moreover associated with each
positive root $\a\in\Delta_+$, there are a raising operator
$E_{\a}$ which altogether spans the subalgebra
$\lt(osp(2n|2n)\rt)_+$, a lowering operator $F_{\a}$ which
altogether spans the subalgebra $\lt(osp(2n|2n)\rt)_-$  and a
Cartan generator $H_{\a}$ which altogether spans the Cartan
subalgebra $\textbf{h}$. Then one has the Cartan-Weyl
decomposition of $osp(2n|2n)$ \bea
 osp(2n|2n)=\lt(osp(2n|2n)\rt)_{-}\oplus\textbf{h}
 \oplus \lt(osp(2n|2n)\rt)_+.\no
\eea

Hereafter, we adopt the convention that \bea
 E_i\equiv E_{\a_i},\quad F_i\equiv F_{\a_i}, \qquad
 i=1,\ldots,2n.\label{convent}
\eea We remark that the $\Zb_2$-grading of the generators
associated with the simple roots and the Cartan subalgebra are:
\bea
  &&[E_n]=[F_n]=1,\qquad [E_i]=[F_i]=0,\,{\rm for}\, i\neq n,\\
  &&[g]=0,\qquad \forall g\in\textbf{h}.
\eea The matrix realization of the generators associated with all
roots of $osp(2n|2n)$ is given in Appendix A, from which one may
derive the structure constants $f_{ij}^m$ of the algebra in
(\ref{construct-constant}) for the particular choice of the basis.


\section{Free-field realization of the $osp(2n|2n)$ currents}
 \label{DIR} \setcounter{equation}{0}

\subsection{Differential operator realization of $osp(2n|2n)$}

Let us introduce a bosonic coordinate ($x_{m,l}$, $\bar{x}_{m,l}$,
$x_l$, $y_{i,j}$ or $\bar{y}_{i,j}$ for $m<l$ and $i<j$) with a
$\Zb_2$-grading zero: $[x]=[\bar{x}]=[y]=[\bar{y}]=0$ associated
with each positive even root (resp. $\d_m-\d_l$, $\d_m+\d_l$,
$2\d_l$, $\e_i-\e_j$ or $\e_i+\e_j$ for $m<l$ and $i<j$), and a
fermionic coordinate ($\theta_{l,i}$ or $\bar{\theta}_{l,i}$) with
a $\Zb_2$-grading one: $[\theta]=[\bar{\theta}]=1$ associated with
each positive odd root (resp. $\d_l-\e_i$ or $\d_l+\e_i$). These
coordinates satisfy the following (anti)commutation relations:
\bea
 &&[x_{i,j},x_{m,l}]=0,\,\,[\partial_{x_{i,j}},\partial_{x_{m,l}}]=0,
   \,\,[\partial_{x_{i,j}},x_{m,l}]=\d_{im}\d_{jl},\label{Fundament-Comm-1}\\
 &&[\bar{x}_{i,j},\bar{x}_{m,l}]=0,\,\,
   [\partial_{\bar{x}_{i,j}},\partial_{\bar{x}_{m,l}}]=0, \,\,
   [\partial_{\bar{x}_{i,j}},\bar{x}_{m,l}]=\d_{im}\d_{jl},\\
 &&[x_m,x_{l}]=0,\,\,[\partial_{x_m},\partial_{x_{l}}]=0,
   \,\,[\partial_{x_{m}},x_{l}]=\d_{ml},\\
 &&[y_{i,j},y_{m,l}]=0,\,\,[\partial_{y_{i,j}},\partial_{y_{m,l}}]=0,
   \,\,[\partial_{y_{i,j}},y_{m,l}]=\d_{im}\d_{jl},\\
 &&[\bar{y}_{i,j},\bar{y}_{m,l}]=0,\,\,
   [\partial_{\bar{y}_{i,j}},\partial_{\bar{y}_{m,l}}]=0, \,\,
   [\partial_{\bar{y}_{i,j}},\bar{y}_{m,l}]=\d_{im}\d_{jl},\\
 &&[\theta_{i,j},\theta_{m,l}]=0,\,\,[\partial_{\theta_{i,j}},\partial_{\theta_{m,l}}]=0,
   \,\,[\partial_{\theta_{i,j}},\theta_{m,l}]=\d_{im}\d_{jl},\\
 &&[\bar{\theta}_{i,j},\bar{\theta}_{m,l}]=0,\,\,
   [\partial_{\bar{\theta}_{i,j}},\partial_{\bar{\theta}_{m,l}}]=0, \,\,
   [\partial_{\bar{\theta}_{i,j}},\bar{\theta}_{m,l}]=\d_{im}\d_{jl},
 \label{Fundament-Comm-2}
\eea and the other (anti)commutation relations are vanishing. Let
$\langle\L|$ be the highest weight vector of the representation of
$osp(2n|2n)$ with highest weights $\{\l_i\}$ , satisfying the
following conditions: \bea
 &&\langle\L|F_i=0,\qquad\qquad 1\leq i\leq 2n,\label{highestweight-1}\\
 &&\langle\L|H_i=\l_i\,\langle\L|,\qquad\qquad 1\leq i\leq
  2n.\label{Lowestweight-2}
\eea Here the generators $H_i$ are expressed in terms of some
linear combinations of $H_{\a}$ (\ref{D-H-1})-(\ref{D-H-2}). An
arbitrary vector in the corresponding Verma module  is
parametrized by $\langle\L|$ and the corresponding bosonic and
fermionic coordinates as
\bea
 \langle\L;x,\bar{x};y,\bar{y};\theta,\bar{\theta}|=
 \langle\L|G_{+}(x,\bar{x};y,\bar{y};\theta,\bar{\theta}),\label{States-D}
\eea where $G_{+}(x,\bar{x};y,\bar{y};\theta,\bar{\theta})$ is
given by (c.f. \cite{Ito90,Ras98}\footnote{It practically would be
very {\it involved} (if not impossible) to apply the general
procedure proposed in \cite{Ito90,Ras98} for $osp(2n|2n)$ with a
large $n$, however, our choice of
$G_{+}(x,\bar{x};y,\bar{y};\theta,\bar{\theta})$
(\ref{G-1})-(\ref{G-2}) allow us to obtain the explicit
expressions of the differential operator realization of the
algebra with a generic $n$ (see (\ref{Diff-D-1})-(\ref{Diff-D-2})
below).}) \bea
   G_{+}(x,\bar{x};y,\bar{y};\theta,\bar{\theta})&=&\lt(\bar{G}_{2n-1,2n}\,G_{2n-1,2n}\rt)\ldots
      \lt(\bar{G}_{n+1,n+2}\ldots\bar{G}_{n+1,2n}\,G_{n+1,2n}\ldots G_{n+1,n+2}\rt)\no\\
   &&\quad \times \lt(\bar{G}_{n,n+1}\ldots\bar{G}_{n,2n}
      \,G_{n}\,G_{n,2n}\ldots G_{n,n+1}\rt)\ldots\no\\
   &&\quad \times \lt(\bar{G}_{1,2}\ldots\bar{G}_{1,2n}
      \,G_1\, G_{1,2n}\ldots G_{1,2}\rt).\label{G-1}
\eea Here $G_{i,j}$ and $\bar{G}_{i,j}$ are given by
\bea
  &&G_{m,l}=e^{x_{m,l}E_{\d_m-\d_l}},\quad
           \bar{G}_{m,l}=e^{\bar{x}_{m,l}E_{\d_m+\d_l}},\qquad 1\leq m<l\leq n,\\
  &&G_{l}=e^{x_lE_{2\d_l}},\quad G_{l,n+i}=e^{\theta_{l,i}E_{\d_l-\e_i}},\quad
           \bar{G}_{l,n+i}=e^{\bar{\theta}_{l,i}E_{\d_l+\e_i}},\qquad 1\leq l,i\leq n,\\
  &&G_{n+i,n+j}=e^{y_{i,j}E_{\e_i-\e_j}},\qquad
           \bar{G}_{n+i,n+j}=e^{\bar{y}_{i,j}E_{\e_i+\e_j}},\qquad 1\leq i<j\leq
           n.\label{G-2}
\eea One can define a differential operator realization
$\rho^{(d)}$ of the generators of $osp(2n|2n)$ by
\bea
 \rho^{(d)}(g)\,\langle\L;x,\bar{x};y,\bar{y};\theta,\bar{\theta}|
    \equiv \langle\L;x,\bar{x};y,\bar{y};\theta,\bar{\theta}|\,g,\qquad
     \forall g\in osp(2n|2n).\label{definition-D}
\eea Here $\rho^{(d)}(g)$ is a differential operator of the
coordinates $\{x,\,\bar{x};y,\bar{y};\theta,\bar{\theta}\}$
associated with the generator $g$, which can be obtained from the
defining relation (\ref{definition-D}). The defining relation also
assures that  the differential operator realization is actually a
representation of $osp(2n|2n)$. Therefore it is sufficient to give
the differential operators related to the simple roots, as the
others can be constructed through the simple ones by the
(anti)commutation relations. Using the relation
(\ref{definition-D}) and the Baker-Campbell-Hausdorff formula,
after some algebraic manipulations, we obtain the following
differential operator representation of the simple generators:
\bea
  \rho^{(d)}(E_l)&=&\sum_{m=1}^{l-1}
    \lt(x_{m,l}\partial_{x_{m,l+1}}-\bar{x}_{m,l+1}\partial_{\bar{x}_{m,l}}\rt)
    +\partial_{x_{l,l+1}},\qquad 1\leq l\leq n-1,\label{Diff-D-1}\\
  \rho^{(d)}(E_n)&=&\sum_{m=1}^{n-1}
    \lt(x_{m,n}\partial_{\theta_{m,1}}+\bar{\theta}_{m,1}
    \partial_{\bar{x}_{m,n}}\rt)+\partial_{\theta_{n,1}},\\
  \rho^{(d)}(E_{n+i})&=&\sum_{m=1}^n\lt(\theta_{m,i}\partial_{\theta_{m,i+1}}
    -\bar{\theta}_{m,i+1}\partial_{\bar{\theta}_{m,i}}\rt)\no\\
    &&+\sum_{m=1}^{i-1}
    \lt(y_{m,i}\partial_{y_{m,i+1}}-\bar{y}_{m,i+1}\partial_{\bar{y}_{m,i}}\rt)
    +\partial_{y_{i,i+1}},\qquad 1\leq i\leq n-1,\\
  \rho^{(d)}(E_{2n})&=&\sum_{m=1}^n\lt(2\theta_{m,n-1}\theta_{m,n}\partial_{x_m}+
    \theta_{m,n-1}\partial_{\bar{\theta}_{m,n}}
    -\theta_{m,n}\partial_{\bar{\theta}_{m,n-1}}\rt)\no\\
    &&+\sum_{m=1}^{n-2}
    \lt(y_{m,n-1}\partial_{\bar{y}_{m,n}}-y_{m,n}\partial_{\bar{y}_{m,n-1}}\rt)
    +\partial_{\bar{y}_{n-1,n}},\\[8pt]
  \rho^{(d)}(F_l)&=&\sum_{m=1}^{l-1}\lt(x_{m,l+1}\partial_{x_{m,l}}
    -\bar{x}_{m,l}\partial_{\bar{x}_{m,l+1}}\rt)
    -x_l\partial_{\bar{x}_{l,l+1}}-2\bar{x}_{l,l+1}\partial_{x_{l+1}}\no\\
    &&+\sum_{m=l+2}^n\lt(x_{l,m}\bar{x}_{l,m}\partial_{\bar{x}_{l,l+1}}
    -x_{l,m}\partial_{x_{l+1,m}}
    -2\bar{x}_{l,m}x_{l+1,m}\partial_{x_{l+1}}
    -\bar{x}_{l,m}\partial_{\bar{x}_{l+1,m}}\rt)\no\\
    &&-\sum_{m=1}^n\lt(\theta_{l,m}\bar{\theta}_{l,m}\partial_{\bar{x}_{l,l+1}}
    +\theta_{l,m}\partial_{\theta_{l+1,m}}
    +2\bar{\theta}_{l,m}\theta_{l+1,m}\partial_{x_{l+1}}
    +\bar{\theta}_{l,m}\partial_{\bar{\theta}_{l+1,m}}\rt)\no\\
    &&-x_{l,l+1}^2\partial_{x_{l,l+1}}+2x_{l,l+1}x_{l+1}\partial_{x_{l+1}}
    -2x_{l,l+1}x_l\partial_{x_l}\no\\
    &&+x_{l,l+1}\hspace{-0.1truecm}\lt[
    \sum_{m=l+2}^n
    \hspace{-0.1truecm}\lt(x_{l+1,m}\partial_{x_{l+1,m}}
    \hspace{-0.1truecm}+\hspace{-0.1truecm}
    \bar{x}_{l+1,m}\partial_{\bar{x}_{l+1,m}}
    \hspace{-0.1truecm}-\hspace{-0.1truecm}x_{l,m}\partial_{x_{l,m}}
    -\bar{x}_{l,m}\partial_{\bar{x}_{l,m}}\rt)\rt]\no\\
    &&+x_{l,l+1}\lt[\sum_{m=1}^n\lt(\theta_{l+1,m}\partial_{\theta_{l+1,m}}
    +\bar{\theta}_{l+1,m}\partial_{\bar{\theta}_{l+1,m}}
    -\theta_{l,m}\partial_{\theta_{l,m}}
    -\bar{\theta}_{l,m}\partial_{\bar{\theta}_{l,m}}\rt)\rt]\no\\
    &&+x_{l,l+1}(\l_l-\l_{l+1}),\qquad\qquad 1\leq l\leq
    n-1,\\[6pt]
  \rho^{(d)}(F_n)&=&\sum_{m=1}^{n-1}\lt(\theta_{m,1}\partial_{x_{m,n}}
    -\bar{x}_{m,n}\partial_{\bar{\theta}_{m,1}}\rt)-x_n\partial_{\bar{\theta}_{n,1}}\no\\
    &&+\sum_{m=2}^n\lt(\theta_{n,m}\partial_{y_{1,m}}
    -\theta_{n,m}\bar{\theta}_{n,m}\partial_{\bar{\theta}_{n,1}}
    +\bar{\theta}_{n,m}\partial_{\bar{y}_{1,m}}\rt)\no\\
    &&-\theta_{n,1}\sum_{m=2}^n\lt(\theta_{n,m}\partial_{\theta_{n,m}}
    +\bar{\theta}_{n,m}\partial_{\bar{\theta}_{n,m}}+y_{1,m}\partial_{y_{1,m}}
    +\bar{y}_{1,m}\partial_{\bar{y}_{1,m}}\rt)\no\\
    &&-2\theta_{n,1}x_n\partial_{x_n}
    -2\theta_{n,1}\bar{\theta}_{n,1}\partial_{\bar{\theta}_{n,1}}
    +\theta_{n,1}(\l_{n}+\l_{n+1}),\\
 \rho^{(d)}(F_{n+i})&=&\sum_{m=1}^n(\theta_{m,i+1}\partial_{\theta_{m,i}}
    -\bar{\theta}_{m,i}\partial_{\bar{\theta}_{m,i+1}})
    +\sum_{m=1}^{i-1}(y_{m,i+1}\partial_{y_{m,i}}-\bar{y}_{m,i}\partial_{\bar{y}_{m,i+1}})\no\\
    &&+\sum_{m=i+2}^n\lt(y_{i,m}\bar{y}_{i,m}\partial_{\bar{y}_{i,i+1}}
    -y_{i,m}\partial_{y_{i+1,m}}-\bar{y}_{i,m}\partial_{\bar{y}_{i+1,m}}\rt)\no\\
    &&+y_{i,i+1}\sum_{m=i+2}^n\lt(y_{i+1,m}\partial_{y_{i+1,m}}
    +\bar{y}_{i+1,m}\partial_{\bar{y}_{i+1,m}}
    -y_{i,m}\partial_{y_{i,m}}
    -\bar{y}_{i,m}\partial_{\bar{y}_{i,m}}\rt)\no\\
    &&-y^2_{i,i+1}\partial_{y_{i,i+1}}+y_{i,i+1}(\l_{n+i}-\l_{n+i+1}),
    \qquad 1\leq i\leq n-1,\\
  \rho^{(d)}(F_{2n})&=&\sum_{m=1}^n\lt(\bar{\theta}_{m,n}\partial_{\theta_{m,n-1}}
    +2\bar{\theta}_{m,n-1}\bar{\theta}_{m,n}\partial_{x_m}
    -\bar{\theta}_{m,n-1}\partial_{\theta_{m,n}}\rt)\no\\
    &&+\sum_{m=1}^{n-2}\lt(\bar{y}_{m,n}\partial_{y_{m,n-1}}-\bar{y}_{m,n-1}\partial_{y_{m,n}} \rt)
    -\bar{y}^2_{n-1,n}\partial_{\bar{y}_{n-1,n}}\no\\
    &&+\bar{y}_{n-1,n}(\l_{2n-1}+\l_{2n}),\\[8pt]
  \rho^{(d)}(H_l)&=&\sum_{m=1}^{l-1}\lt(x_{m,l}\partial_{x_{m,l}}
     -\bar{x}_{m,l}\partial_{\bar{x}_{m,l}}\rt)
     -\sum_{m=l+1}^{n}\lt(x_{l,m}\partial_{x_{l,m}}
     +\bar{x}_{l,m}\partial_{\bar{x}_{l,m}}\rt)\no\\
     &&-\sum_{m=1}^n\lt(\theta_{l,m}\partial_{\theta_{l,m}}
     +\bar{\theta}_{l,m}\partial_{\bar{\theta}_{l,m}}\rt)
     -2x_l\partial_{x_l}+\l_l,\qquad 1\leq l\leq n,\\
  \rho^{(d)}(H_{n+i})&=&\sum_{m=1}^n\lt( \theta_{m,i}\partial_{\theta_{m,i}}
     -\bar{\theta}_{m,i}\partial_{\bar{\theta}_{m,i}}\rt)
     +\sum_{m=1}^{i-1}\lt(y_{m,i}\partial_{y_{m,i}}
     -\bar{y}_{m,i}\partial_{\bar{y}_{m,i}}\rt)\no\\
     &&-\sum_{m=i+1}^n\lt(y_{i,m}\partial_{y_{i,m}}
     +\bar{y}_{i,m}\partial_{\bar{y}_{i,m}}\rt)+\l_{n+i},
     \qquad 1\leq i \leq n.\label{Diff-D-2}
\eea

A direct computation shows that these differential operators
(\ref{Diff-D-1})-(\ref{Diff-D-2}) satisfy the $osp(2n|2n)$
(anti)commutation relations corresponding to the simple roots and
the associated Serre relations. This implies that the differential
representation of non-simple generators can be consistently
constructed from the simple ones. Hence, we have obtained an
explicit differential realization of $osp(2n|2n)$.

\subsection{Free-field realization of $osp(2n|2n)_k$}

With the help of the differential realization given by
(\ref{Diff-D-1})-(\ref{Diff-D-2}) we can construct the explicit
free-field representation of the $osp(2n|2n)$ current algebra at
arbitrary level $k$ in terms of $n\times (2n-1)$ bosonic $\b$-$\g$
pairs $\{(\b_{i,j},\,\g_{i,j}),\,
(\bar{\b}_{i,j}\bar{\g}_{i,j}),\,(\b_i,\g_i),\,(\b'_{i,j},\,\g'_{i,j}),\,
(\bar{\b}'_{i,j}\bar{\g}'_{i,j}),\, 1\leq i<j\leq n\}$, $2n^2$
ferminic $b-c$ pairs
$\{(\Psi^+_{i,j},\Psi_{i,j}),\,(\bar{\Psi}^+_{i,j},\bar{\Psi}_{i,j}),\,1\leq
i,j\leq n\}$ and $2n$ free scalar fields $\phi_i$,
$i=1,\ldots,2n$. These free fields obey the following OPEs: \bea
  &&\hspace{-0.8truecm}\b_{i,j}(z)\,\g_{m,l}(w)=-\g_{m,l}(z)\,\b_{i,j}(w)=
      \frac{\d_{im}\d_{jl}}{(z-w)},\,\,1\leq i<j\leq n,\,\,1\leq
      m<l\leq n,\label{OPE-F-1}\\
  &&\hspace{-0.8truecm}\bar{\b}_{i,j}(z)\,\bar{\g}_{m,l}(w)=-\bar{\g}_{m,l}(z)\,
      \bar{\b}_{i,j}(w)= \frac{\d_{im}\d_{jl}}{(z-w)},\,\,1\leq i<j\leq
      n,\,\,1\leq m<l\leq n,\\
  &&\hspace{-0.8truecm}\b_{m}(z)\,\g_{l}(w)=-\g_{m}(z)\,\b_{l}(w)=
      \frac{\d_{ml}}{(z-w)},\,\,1\leq m,l\leq n,\\
  &&\hspace{-0.8truecm}\b'_{i,j}(z)\,\g'_{m,l}(w)=-\g'_{m,l}(z)\,\b'_{i,j}(w)=
      \frac{\d_{im}\d_{jl}}{(z-w)},\,\,1\leq i<j\leq n,\,\,1\leq
      m<l\leq n,\\
  &&\hspace{-0.8truecm}\bar{\b}'_{i,j}(z)\,\bar{\g}'_{m,l}(w)=-\bar{\g}'_{m,l}(z)\,
      \bar{\b}'_{i,j}(w)= \frac{\d_{im}\d_{jl}}{(z-w)},\,\,1\leq i<j\leq
      n,\,\,1\leq m<l\leq n,\\
  &&\hspace{-0.8truecm}\Psi^+_{m,i}(z)\,\Psi_{l,j}(w)=\Psi_{l,j}(z)\,
      \Psi^+_{m,i}(w)= \frac{\d_{ml}\d_{ij}}{(z-w)},\,\,m,l,i,j=1,\ldots,n,\\
  &&\hspace{-0.8truecm}\bar{\Psi}^+_{m,i}(z)\,\bar{\Psi}_{l,j}(w)=\bar{\Psi}_{l,j}(z)\,
      \bar{\Psi}^+_{m,i}(w)= \frac{\d_{ml}\d_{ij}}{(z-w)},\,\,m,l,i,j=1,\ldots,n,\\
  &&\hspace{-0.8truecm}\phi_m(z)\phi_l(w)=-\d_{ml}\,
      \ln(z-w),\,\,\,\,\,1\leq m,l\leq n,\\
  &&\hspace{-0.8truecm}\phi_{n+i}(z)\phi_{n+j}(w)=\d_{ij}\,
      \ln(z-w),\,\,\,\,\,1\leq i,j\leq n,\label{OPE-F-2}
\eea and the other OPEs are trivial.

The free-field realization of the $osp(2n|2n)$ current algebra is
obtained by the  substitution in the differential realization
(\ref{Diff-D-1})-(\ref{Diff-D-2}) of $osp(2n|2n)$, \bea
 &&x_{m,l}\longrightarrow \g_{m,l}(z),\quad \partial_{x_{m,l}}
   \longrightarrow \b_{m,l}(z),\quad 1\leq m<l\leq n,\label{sub-1}\\
 &&\bar{x}_{m,l}\longrightarrow \bar{\g}_{m,l}(z),\quad
   \partial_{\bar{x}_{m,l}} \longrightarrow \bar{\b}_{m,l}(z),
   \quad 1\leq m<l\leq n,\\
 &&x_{l}\longrightarrow \g_{l}(z),\quad \partial_{x_{l}}
   \longrightarrow \b_{l}(z),\quad 1\leq l\leq n,\\
 &&y_{i,j}\longrightarrow \g'_{i,j}(z),\quad \partial_{y_{i,j}}
   \longrightarrow \b'_{i,j}(z),\quad 1\leq i<j\leq n,\\
 &&\bar{y}_{i,j}\longrightarrow \bar{\g}'_{i,j}(z),\quad
   \partial_{\bar{y}_{i,j}} \longrightarrow \bar{\b}'_{i,j}(z),
   \quad 1\leq i<j\leq n,\\
 &&\theta_{l,i}\longrightarrow \Psi^+_{l,i}(z),\quad \partial_{\theta_{l,i}}
   \longrightarrow \Psi_{l,i}(z),\quad i,l=1,\ldots,n,\\
 &&\bar{\theta}_{l,i}\longrightarrow \bar{\Psi}^+_{l,i}(z),\quad
   \partial_{\bar{\theta}_{l,i}} \longrightarrow \bar{\Psi}_{l,i}(z),
   \quad i,l=1,\ldots,n,\\
 &&\l_j\longrightarrow \sqrt{k-2}\,\partial\phi_j(z)\qquad
   1\leq j\leq 2n.\label{sub-2}\eea
Moreover, in order that the resulting free-field realization
satisfy the desirable OPEs for $osp(2n|2n)$ currents, one needs to
add certain extra (anomalous) terms which are linear in
$\partial\g(z)$, $\partial\bar{\g}(z)$, $\partial\g'(z)$,
$\partial\bar{\g}'(z)$, $\partial\Psi^+(z)$ and
$\partial\bar{\Psi}^+(z)$ in the expressions of the currents
associated with negative roots (e.g. the last term in the
expressions of $F_i(z)$, see (\ref{Free-D-F-1})-(\ref{Free-D-F-2})
below). Here we present the results for the currents associated
with the simple roots, \bea
  E_{l}(z)&=&\sum_{m=1}^{l-1}\lt(\g_{m,l}(z)\b_{m,l+1}(z)-
    \bar{\g}_{m,l+1}(z)\bar{\b}_{m,l}(z)\rt)+\b_{l,l+1}(z),\quad
    1\leq l\leq n-1,\label{Free-D-1}\\
  E_{n}(z)&=&\sum_{m=1}^{n-1}\lt(\g_{m,n}(z)\Psi_{m,1}(z)+
    \bar{\Psi}^+_{m,1}(z)\bar{\b}_{m,n}(z)\rt)+\Psi_{n,1}(z),\\
  E_{n+i}(z)&=&\sum_{m=1}^n\lt(\Psi^+_{m,i}(z)\Psi_{m,i+1}(z)
    -\bar{\Psi}^+_{m,i+1}(z)\bar{\Psi}_{m,i}(z)\rt)\no\\
    &&+\sum_{m=1}^{i-1}\lt(\g'_{m,i}(z)\b'_{m,i+1}(z)
    \hspace{-0.1truecm}-\hspace{-0.1truecm}
    \bar{\g}'_{m,i+1}(z)\bar{\b}'_{m,i}(z)\rt)
    \hspace{-0.1truecm}+\hspace{-0.1truecm}\b'_{i,i+1}(z),
    \quad 1\leq i\leq n-1,\\
  E_{2n}(z)&=&\sum_{m=1}^n\hspace{-0.1truecm}
    \lt(2\Psi^+_{m,n-1}(z)\Psi^+_{m,n}(z)
    \b_m(z)\hspace{-0.1truecm}+\hspace{-0.1truecm}
    \Psi^+_{m,n-1}(z)\bar{\Psi}_{m,n}(z)
    \hspace{-0.1truecm}-\hspace{-0.1truecm}
    \Psi^+_{m,n}(z)\bar{\Psi}_{m,n-1}(z)\rt)\no\\
    &&+\sum_{m=1}^{n-2}\lt(\g'_{m,n-1}(z)\bar{\b}'_{m,n}(z)
    -\g'_{m,n}(z)\bar{\b}'_{m,n-1}(z)\rt)+\bar{\b}'_{n-1,n}(z),\\[8pt]
  F_{l}(z)&=&\sum_{m=1}^{l-1}\lt(\g_{m,l+1}(z)\b_{m,l}(z)
    -\bar{\g}_{m,l}(z)\bar{\b}_{m,l+1}(z)\rt)
    -\g_{l}(z)\bar{\b}_{l,l+1}(z)\no\\
    &&-2\bar{\g}_{l,l+1}(z)\b_{l+1}(z)
    +\sum_{m=l+2}^{n}\lt(
    \g_{l,m}(z)\bar{\g}_{l,m}(z)\bar{\b}_{l,l+1}(z)
    -\g_{l,m}(z)\b_{l+1,m}(z)\rt)\no\\
    &&-\sum_{m=l+2}^{n}\lt(
    2\bar{\g}_{l,m}(z)\g_{l+1,m}(z)\b_{l+1}(z)
    +\bar{\g}_{l,m}(z)\bar{\b}_{l+1,m}(z)\rt)\no\\
    &&-\sum_{m=1}^{n}\lt(
    \Psi^+_{l,m}(z)\bar{\Psi}^+_{l,m}(z)\bar{\b}_{l,l+1}(z)
    +\Psi^+_{l,m}(z)\Psi_{l+1,m}(z)\rt)\no\\
    &&-\sum_{m=1}^{n}\lt(
    2\bar{\Psi}^+_{l,m}(z)\Psi^+_{l+1,m}(z)\b_{l+1}(z)
    +\bar{\Psi}^+_{l,m}(z)\bar{\Psi}_{l+1,m}(z)\rt)\no\\
    &&-\g^2_{l,l+1}(z)\b_{l,l+1}(z)-\g_{l,l+1}(z)\sum_{m=l+2}^n
    \lt(\g_{l,m}(z)\b_{l,m}(z)
      +\bar{\g}_{l,m}(z)\bar{\b}_{l,m}(z)\rt)\no\\
    &&+\g_{l,l+1}(z)\sum_{m=l+2}^n\lt(\g_{l+1,m}(z)\b_{l+1,m}(z)+
    \bar{\g}_{l+1,m}(z)\bar{\b}_{l+1,m}(z)\rt)\no\\
    &&-\g_{l,l+1}(z)\sum_{m=1}^n
    \lt(\Psi^+_{l,m}(z)\Psi_{l,m}(z)
      +\bar{\Psi}^+_{l,m}(z)\bar{\Psi}_{l,m}(z)\rt)\no\\
    &&+\g_{l,l+1}(z)\sum_{m=1}^n\lt(\Psi^+_{l+1,m}(z)\Psi_{l+1,m}(z)+
    \bar{\Psi}^+_{l+1,m}(z)\bar{\Psi}_{l+1,m}(z)\rt)\no\\
    &&+2\g_{l,l+1}(z)\g_{l+1}(z)\b_{l+1}(z)
    -2\g_{l,l+1}(z)\g_{l}(z)\b_{l}(z)\no\\
    &&+\sqrt{k-2}\g_{l,l+1}(z)\lt(\partial\phi_l(z)\hspace{-0.1truecm}
    -\hspace{-0.1truecm}\partial\phi_{l+1}(z)\rt)\no\\
    &&+(-k+2(l-1))\partial\g_{l,l+1}(z),
    \quad 1\leq l\leq n-1,\label{Free-D-F-1}\\
  F_n(z)&=&\sum_{m=1}^{n-1}\lt(\Psi^+_{m,1}(z)\b_{m,n}(z)
     -\bar{\g}_{m,n}(z)\bar{\Psi}_{m,1}(z)\rt)
     -\g_{n}(z)\bar{\Psi}_{n,1}(z)\no\\
     &&+\sum_{m=2}^n\lt(\Psi^+_{n,m}(z)\b'_{1,m}(z)
     -\Psi^+_{n,m}(z)\bar{\Psi}^+_{n,m}(z)\bar{\Psi}_{n,1}(z)
     +\bar{\Psi}^+_{n,m}(z)\bar{\b}'_{1,m}(z)\rt)\no\\
     &&-\hspace{-0.1truecm}\Psi^+_{n,1}(z)\hspace{-0.1truecm}\sum_{m=2}^n
     \hspace{-0.1truecm}\lt(\Psi^+_{n,m}(z)\Psi_{n,m}(z)
     \hspace{-0.1truecm}+\hspace{-0.1truecm}
     \bar{\Psi}^+_{n,m}(z)\bar{\Psi}_{n,m}(z)\rt)
     \hspace{-0.1truecm}-\hspace{-0.1truecm}
     2\Psi^+_{n,1}(z)\bar{\Psi}^+_{n,1}(z)\bar{\Psi}_{n,1}(z)\no\\
     &&-\Psi^+_{n,1}(z)\sum_{m=2}^n\lt(\g'_{1,m}(z)\b'_{1,m}(z)
     +\bar{\g}'_{1,m}(z)\bar{\b}'_{1,m}(z)\rt)
     -2\Psi^+_{n,1}(z)\g_n(z)\b_n(z)\no\\
     &&+\sqrt{k-2}\Psi^+_{n,1}(z)\lt(
     \partial\phi_{n}(z)\hspace{-0.1truecm}+\hspace{-0.1truecm}\partial\phi_{n+1}(z)\rt)\no\\
     &&+(-k+2(n-1))\partial\,\Psi^+_{n,1}(z),\\
  F_{n+i}(z)&=&\sum_{m=1}^n\lt(\Psi^+_{m,i+1}(z)\Psi_{m,i}(z)
     -\bar{\Psi}^+_{m,i}(z)\bar{\Psi}_{m,i+1}(z)\rt)\no\\
     &&+\sum_{m=1}^{i-1}\lt(\g'_{m,i+1}(z)\b'_{m,i}(z)
     -\bar{\g}'_{m,i}(z)\bar{\b}'_{m,i+1}(z)\rt)\no\\
     &&+\sum_{m=i+2}^n\lt(\g'_{i,m}(z)\bar{\g}'_{i,m}(z)\bar{\b}'_{i,i+1}(z)
     -\g'_{i,m}(z)\b'_{i+1,m}(z)
     -\bar{\g}'_{i,m}(z)\bar{\b}'_{i+1,m}(z)\rt)\no\\
     &&+\g'_{i,i+1}(z)\sum_{m=i+2}^n\lt(\g'_{i+1,m}(z)\b'_{i+1,m}(z)
     +\bar{\g}'_{i+1,m}(z)\bar{\b}'_{i+1,m}(z)\rt)\no\\
     &&-\g'_{i,i+1}(z)\sum_{m=i+2}^n\lt(\g'_{i,m}(z)\b'_{i,m}(z)
     +\bar{\g}'_{i,m}(z)\bar{\b}'_{i,m}(z)\rt)\no\\
     &&-\g'_{i,i+1}(z)\g'_{i,i+1}(z)\b'_{i,i+1}(z)
     +\sqrt{k-2}\g'_{i,i+1}(z)\lt(\partial\phi_{n+i}(z)
     -\partial\phi_{n+i+1}(z)\rt)\no\\
     &&+\lt(k+2(i-n-1)\rt)\partial \g'_{i,i+1}(z),\qquad 1\leq i\leq n-1,\\[8pt]
  F_{2n}(z)&=&\sum_{m=1}^n\hspace{-0.1truecm}\lt(\bar{\Psi}^+_{m,n}(z)\Psi_{m,n-1}(z)
     \hspace{-0.1truecm}+\hspace{-0.1truecm}
     2\bar{\Psi}^+_{m,n-1}(z)\bar{\Psi}^+_{m,n}(z)\b_m(z)
     \hspace{-0.1truecm}-\hspace{-0.1truecm}\bar{\Psi}^+_{m,n-1}(z)\Psi_{m,n}(z)\rt)\no\\
     &&+\sum_{m=1}^{n-2}\lt(\bar{\g}'_{m,n}(z)\b'_{m,n-1}(z)
     -\bar{\g}'_{m,n-1}(z)\b'_{m,n}(z)\rt)\no\\
     &&-\bar{\g}'_{n-1,n}(z)\bar{\g}'_{n-1,n}(z)\bar{\b}'_{n-1,n}(z)\no\\
     &&+\sqrt{k-2}\bar{\g}'_{n-1,n}(z)\lt(\partial\phi_{2n-1}(z)
     +\partial\phi_{2n}(z)\rt)+\lt(k-4\rt)\partial\bar{\g}'_{n-1,n}(z),\label{Free-D-F-2}\\[8pt]
  H_l(z)&=&\hspace{-0.16truecm}\sum_{m=1}^{l-1}\hspace{-0.16truecm}\lt(\g_{m,l}(z)\b_{m,l}(z)
     \hspace{-0.1truecm}-\hspace{-0.1truecm}\bar{\g}_{m,l}(z)\bar{\b}_{m,l}(z)\rt)
     \hspace{-0.1truecm}-\hspace{-0.1truecm}
     \sum_{m=l+1}^{n}\hspace{-0.16truecm}\lt(\g_{l,m}(z)\b_{l,m}(z)\hspace{-0.1truecm}
     +\hspace{-0.1truecm}\bar{\g}_{l,m}(z)\bar{\b}_{l,m}(z)\rt)\no\\
     &&-2\g_l(z)\b_{l}(z)-\sum_{m=1}^n\lt(\Psi^+_{l,m}(z)\Psi_{l,m}(z)
     +\bar{\Psi}^+_{l,m}(z)\bar{\Psi}_{l,m}(z)\rt)\no\\
     &&+\sqrt{k-2}\partial\phi_l(z),\qquad\qquad 1\leq l\leq
     n,\\
  H_{n+i}(z)&=&\hspace{-0.16truecm}\sum_{m=1}^{n}\hspace{-0.16truecm}
     \lt(\Psi^+_{m,i}(z)\Psi_{m,i}(z)
     \hspace{-0.1truecm}-\hspace{-0.1truecm}\bar{\Psi}^+_{m,i}(z)\bar{\Psi}_{m,i}(z)\rt)
     \hspace{-0.1truecm}+\hspace{-0.1truecm}
     \sum_{m=1}^{i-1}\hspace{-0.16truecm}\lt(\g'_{m,i}(z)\b'_{m,i}(z)\hspace{-0.1truecm}
     -\hspace{-0.1truecm}\bar{\g}'_{m,i}(z)\bar{\b}'_{m,i}(z)\rt)\no\\
     &&-\sum_{m=i+1}^n\lt(\g'_{i,m}(z)\b'_{i,m}(z)
     +\bar{\g}'_{i,m}(z)\bar{\b}'_{i,m}(z)\rt)\no\\
     &&+\sqrt{k-2}\partial\phi_{n+i}(z),\qquad\qquad 1\leq i\leq
     n.\label{Free-D-2}
\eea Here and throughout normal ordering of free-fields is implied
whenever necessary. The free-field realization of the currents
associated with the non-simple roots can be obtained from the OPEs
of the simple ones. We can straightforwardly check that the above
free-field realization of the currents satisfies the OPEs of the
$osp(2n|2n)$ current algebra: Direct calculation shows that there
are at most second order singularities (e.g. $1\over(z-w)^{2}$) in
the OPEs of the currents. Comparing with the definition of the
current algebra (\ref{current-OPE}), terms with first order
singularity (e.g. the coefficients of $1\over(z-w)$) are fulfilled
due to the very substitution (\ref{sub-1})-(\ref{sub-2}) and the
fact that the differential operator realizations (\ref{Diff-D-1})-
(\ref{Diff-D-2}) are a representation of the corresponding
finite-dimensional   superalgebra $osp(2n|2n)$; terms with second
order singularity $1\over(z-w)^{2}$  also match those in the
definition (\ref{current-OPE}) after the suitable choice we made
for the anomalous terms in the expressions of the currents
associated with negative roots.

The free-field realization of the $osp(2n|2n)$ current algebra
(\ref{Free-D-1})-(\ref{Free-D-2}) gives rise to  the Fock
representations of the current algebra in terms of the free
fields (\ref{OPE-F-1})-(\ref{OPE-F-2}). These representations are
in general not irreducible for the current algebra. In order to
obtain irreducible ones, one needs certain screening charges,
which are the integrals of screening currents (see
(\ref{Screen-D-1})-(\ref{Screen-D-2}) below), and performs the
cohomology procedure as in  \cite{Fat86,Fei90,Bou90,Ber89}. We
shall construct the associated screening currents in section 5.


\section{Energy-momentum tensor} \label{EMT}
\setcounter{equation}{0}

In this section we construct the free-field realization of the
Sugawara energy-momentum tensor $T(z)$ of the $osp(2n|2n)$ current
algebra. After a tedious calculation, we find \bea
  T(z)&=&\frac{1}{2\lt(k-2\rt)}\lt\{-\sum_{m<l}\lt(
     E_{\d_m-\d_l}(z)F_{\d_m-\d_l}(z)+F_{\d_m-\d_l}(z)E_{\d_m-\d_l}(z)\rt)\rt.\no\\
  &&-\sum_{m<l}\lt(
     E_{\d_m+\d_l}(z)F_{\d_m+\d_l}(z)+F_{\d_m+\d_l}(z)E_{\d_m+\d_l}(z)\rt)\no\\
  &&-\sum_{l=1}^n\lt[\,2\lt(
     E_{2\d_l}(z)F_{2\d_l}(z)+F_{2\d_l}(z)E_{2\d_l}(z)\rt)+H_l(z)H_l(z)\,\rt]\no\\
  &&+\sum_{l=1}^n\sum_{i=1}^n\lt(
     E_{\d_l-\e_i}(z)F_{\d_l-\e_i}(z)-F_{\d_l-\e_i}(z)E_{\d_l-\e_i}(z)\rt)\no\\
  &&+\sum_{l=1}^n\sum_{i=1}^n\lt(
     E_{\d_l+\e_i}(z)F_{\d_l+\e_i}(z)-F_{\d_l+\e_i}(z)E_{\d_l+\e_i}(z)\rt)\no\\
  &&+\sum_{i<j}\lt(E_{\e_i-\e_j}(z)F_{\e_i-\e_j}(z)
     +F_{\e_i-\e_j}(z)E_{\e_i-\e_j}(z)\rt)\no\\
  &&+\lt.\sum_{i<j}\lt(E_{\e_i+\e_j}(z)F_{\e_i+\e_j}(z)+F_{\e_i+\e_j}(z)E_{\e_i+\e_j}(z)\rt)
     +\sum_{i=1}^nH_{n+i}(z)H_{n+i}(z)\rt\}\no\\
  &=&-\sum_{l=1}^n\lt(\frac{1}{2}\partial\phi_l(z)\partial\phi_l(z)-
      \frac{1-l}{\sqrt{k-2}}\partial^2\phi_l(z)\rt)\no\\
  &&+\sum_{i=1}^n\lt(\frac{1}{2}\partial\phi_{n+i}(z)\partial\phi_{n+i}(z)-
      \frac{n-i}{\sqrt{k-2}}\partial^2\phi_{n+i}(z)\rt)\no\\
  &&+\sum_{m<l}\lt(\b_{m,l}(z)\partial\g_{m,l}(z)+
     \bar{\b}_{m,l}(z)\partial\bar{\g}_{m,l}(z)\rt)
     +\sum_{l=1}^n\b_l(z)\partial\g_l(z)\no\\
  &&+\sum_{i<j}\lt(\b'_{i,j}(z)\partial\g'_{i,j}(z)+
     \bar{\b}'_{i,j}(z)\partial\bar{\g}'_{i,j}(z)\rt)\no\\
  &&-\sum_{l=1}^n\sum_{i=1}^n\lt(\Psi_{l,i}(z)\partial\Psi^+_{l,i}(z)
     +\bar{\Psi}_{l,i}(z)\partial\bar{\Psi}^+_{l,i}(z)\rt).\label{Energy-D}
\eea It is straightforward to check that $T(z)$ satisfy the OPE of
the Virasoro algebra, \bea
   T(z)T(w)=\frac{c/2}{(z-w)^4}+\frac{2T(w)}{(z-w)^2}+\frac{\partial
        T(w)}{(z-w)},
\eea with a central charge $c=0$. The vanishing central charge of
the energy-momentum tensor $T(z)$ (\ref{Energy-D}) is a simple
consequence of the fact that the superdimension of $osp(2n|2n)$ is
zero. Moreover, we find that with regard to the energy-momentum
tensor $T(z)$ defined by (\ref{Energy-D}) the $osp(2n|2n)$
currents associated with the simple roots
(\ref{Free-D-1})-(\ref{Free-D-2}) are indeed primary fields with
conformal dimensional one, namely, \bea
  T(z)E_{i}(w)&=&\frac{E_{i}(w)}{(z-w)^2}+\frac{\partial
    E_{i}(w)}{(z-w)},\,\,1\leq i\leq 2n,\no\\
  T(z)F_{i}(w)&=&\frac{F_{i}(w)}{(z-w)^2}+\frac{\partial
    F_{i}(w)}{(z-w)},\,\,1\leq i\leq 2n,\no\\
  T(z)H_{i}(w)&=&\frac{H_{i}(w)}{(z-w)^2}+\frac{\partial
    H_{i}(w)}{(z-w)},\,\,1\leq i\leq 2n.\no
\eea It is expected that  the $osp(2n|2n)$ currents associated
with non-simple roots, which can be constructed through the simple
ones, are also primary fields with conformal dimensional one.
Therefore, $T(z)$ is the very energy-momentum tensor of the
$osp(2n|2n)$ current algebra.


\section{Screening currents} \label{SC}
\setcounter{equation}{0}

Important objects in the application of free-field realizations to
the computation of correlation functions  of the CFTs are
screening currents. A screening current is a primary field with
conformal dimension one and has the property that the singular
part of its OPE with the affine currents is a total derivative.
These properties ensure that the integrated screening currents
(screening charges) may be inserted into correlators while the
conformal or affine Ward identities remain intact
\cite{Dos84,Ber90}.

Free-field realization of the screening currents may be
constructed from certain differential operators \cite{Bou90,Ras98}
which can be defined by the relation, \bea
 \rho^{(d)}\lt(s_{\a}\rt)\,
 \langle\L;x,\bar{x};y,\bar{y};\theta,\bar{\theta}|
 \equiv\langle\L|\,E_{\a}\,
 G_{+}(x,\bar{x};y,\bar{y},\theta,\bar{\theta}),\qquad
 {\rm for}\,\a\in\Delta_+.
 \label{Def-2}
\eea The operators $\rho^{(d)}\lt(s_{\a}\rt)$ ($\a\in\Delta_+$)
give a differential operator realization of the subalgebra
$\lt(osp(2n|2n)\rt)_+$. Again it is sufficient to construct
$s_i\equiv \rho^{(d)}\lt(s_{\a_i}\rt)$ related to the simple
roots. Using (\ref{Def-2}) and the Baker-Campbell-Hausdorff
formula, after some algebraic manipulations, we obtain the
following explicit expressions for $s_i$: \bea
   s_{l}&=&\sum_{m=l+2}^n\hspace{-0.08truecm}\lt(
     -\bar{x}_{l+1,m}x_{l+1,m}\partial_{\bar{x}_{l,l+1}}
     \hspace{-0.08truecm}+\hspace{-0.08truecm}
     \bar{x}_{l+1,m}\partial_{\bar{x}_{l,m}}
     \hspace{-0.08truecm}+\hspace{-0.08truecm}
     2x_{l+1,m}\bar{x}_{l,m}\partial_{x_l}
     \hspace{-0.08truecm}+\hspace{-0.08truecm}
     x_{l+1,m}\partial_{x_{l,m}}\rt)\no\\
     &&+\sum_{m=1}^n\hspace{-0.08truecm}\lt(
     -\bar{\theta}_{l+1,m}\theta_{l+1,m}\partial_{\bar{x}_{l,l+1}}
     \hspace{-0.08truecm}+\hspace{-0.08truecm}
     \bar{\theta}_{l+1,m}\partial_{\bar{\theta}_{l,m}}
     \hspace{-0.08truecm}-\hspace{-0.08truecm}
     2\theta_{l+1,m}\bar{\theta}_{l,m}\partial_{x_l}
     \hspace{-0.08truecm}+\hspace{-0.08truecm}
     \theta_{l+1,m}\partial_{\theta_{l,m}}\rt)\no\\
     &&+x_{l+1}\partial_{\bar{x}_{l,l+1}}
     +2\bar{x}_{l,l+1}\partial_{x_l}+\partial_{x_{l,l+1}},
     \qquad 1\leq l\leq n-1,\label{Scr-P-D-1}\\
   s_n&=&\sum_{m=2}^n\lt(
     \bar{y}_{1,m}\partial_{\bar{\theta}_{n,m}}
     -\bar{y}_{1,m}y_{1,m}\partial_{\bar{\theta}_{n,1}}
     +y_{1,m}\partial_{\theta_{n,m}}
     -2y_{1,m}\bar{\theta}_{n,m}\partial_{x_n}\rt)\no\\
     &&-2\bar{\theta}_{n,1}\partial_{x_n}+\partial_{\theta_{n,1}},\\
   s_{n+i}&=& \sum_{m=i+2}^n\lt(\bar{y}_{i+1,m}\partial_{\bar{y}_{i,m}}
     -\bar{y}_{i+1,m}y_{i+1,m}\partial_{\bar{y}_{i,i+1}}
     +y_{i+1,m}\partial_{y_{i,m}}\rt)\no\\
     &&+\partial_{y_{i,i+1}},\qquad
     1\leq i\leq n-1,\\
   s_{2n}&=&\partial_{\bar{y}_{n-1,n}}.\label{Scr-P-D-2}
\eea {}One may obtain the differential operators $s_{\a}$
associated with the non-simple generators from the above simple
ones. Following the procedure similar to \cite{Bou90,Ras98}, we
find that the free-field realization of the screening currents
$S_i(z)$ corresponding to the differential operators $s_i$ is
given by
 \bea
 S_{l}(z)&=&\lt\{\sum_{m=l+2}^{n}\lt(
    -\bar{\g}_{l+1,m}(z)\g_{l+1,m}(z)\bar{\b}_{l,l+1}(z)
    +\bar{\g}_{l+1,m}(z)\bar{\b}_{l,m}(z)\rt)\rt.\no\\
    &&\quad+\sum_{m=l+2}^{n}\lt(2\g_{l+1,m}(z)\bar{\g}_{l,m}(z)\b_l(z)
    +\g_{l+1,m}(z)\b_{l,m}(z)\rt)+\g_{l+1}(z)\bar{\b}_{l,l+1}(z)\no\\
    &&\quad +2\bar{\g}_{l,l+1}(z)\b_l(z)-\sum_{m=1}^n\lt(
    \bar{\Psi}^+_{l+1,m}(z)\Psi^+_{l+1,m}(z)\bar{\b}_{l,l+1}(z)
    -\bar{\Psi}^+_{l+1,m}(z)\bar{\Psi}_{l,m}(z)\rt)\no\\
    &&\lt.-\sum_{m=1}^n\lt(
    2\Psi^+_{l+1,m}(z)\bar{\Psi}^+_{l,m}(z)\b_{l}(z)
    -\Psi^+_{l+1,m}(z)\Psi_{l,m}(z)\rt)+\b_{l,l+1}(z)\rt\}
    e^{\frac{\a_{l}\cdot\vec{\phi}(z)}{\sqrt{k-2}}},\no\\
    &&\qquad\qquad 1\leq l\leq n-1,\label{Screen-D-1}\\[6pt]
 S_{n}(z)&=&\lt\{\sum_{m=2}^n\lt(\bar{\g}'_{1,m}(z)\bar{\Psi}_{n,m}(z)
    -\bar{\g}'_{1,m}(z)\g'_{1,m}(z)\bar{\Psi}_{n,1}(z)
    -2\g'_{1,m}(z)\bar{\Psi}^+_{n,m}(z)\b_n(z)\rt)\rt.\no\\
    &&\quad\lt.+\sum_{m=2}^n\g'_{1,m}(z)\Psi_{n,m}(z)
    -2\bar{\Psi}^+_{n,1}(z)\b_n(z)+\Psi_{n,1}(z)\rt\}
    e^{\frac{\a_{_n}\cdot\vec{\phi}(z)}{\sqrt{k-2}}}\\
 S_{n+i}(z)&=&\lt\{\sum_{m=i+2}^n\lt(\bar{\g}'_{i+1,m}(z)\bar{\b}'_{i,m}(z)
    -\bar{\g}'_{i+1,m}(z)\g'_{i+1,m}(z)\bar{\b}'_{i,i+1}(z)
    \rt)\rt.\no\\
 &&+\lt.\sum_{m=i+2}^n\g'_{i+1,m}(z)\b'_{i,m}(z)
   +\b'_{i,i+1}(z)\rt\}e^{\frac{\a_{n+i}\cdot\vec{\phi}(z)}{\sqrt{k-2}}},
    \quad 1\leq i\leq n-1,\\
 S_{2n}(z)&=&\bar{\b}'_{n-1,n}(z)e^{\frac{\a_{_{2n}}\cdot\vec{\phi}(z)}{\sqrt{k-2}}}.
 \label{Screen-D-2}
\eea Here $\vec{\phi}(z)$ is
\bea
 \vec{\phi}(z)=\sum_{i=1}^n\lt(\phi_i(z)\,\d_i+\phi_{n+i}(z)\,\e_i\rt).\label{Defin-Phi}
\eea The OPEs of the screening currents with the energy-momentum
tensor and the $osp(2n|2n)$ currents
(\ref{Free-D-1})-(\ref{Free-D-2}) are \bea
  && T(z)S_i(w)=\frac{S_i(w)}{(z-w)^2}+\frac{\partial
       S_i(w)}{(z-w)}=\partial_w\lt\{\frac{S_i(w)}{(z-w)}\rt\},
       \,\,i=1,\ldots,2n,\\
 &&E_{i}(z)S_j(w)=0,\qquad i,j=1\ldots,2n,\\
 &&H_i(z)S_j(w)=0,\qquad i,j=1\ldots,2n,\\
 &&F_i(z)S_j(w)=(-1)^{[[i]]+[F_i]}\d_{ij}\,
\partial_{w}\lt\{\frac{\lt(k-2\rt) \,
e^{\frac{\a_i\cdot\vec{\phi}(w)}{\sqrt{k-2}}}}{(z-w)}\rt\},\no\\
&&\qquad\qquad i,j=1,\ldots,2n.\eea Here $[[i]]$ is given by \bea
 [[i]]=\lt\{\begin{array}{ll}1,&i=1,\ldots,n,\\
 0,&i=n+1,\ldots,2n.\end{array}\rt.\no
\eea The screening currents obtained this way are called screening
currents of the first kind \cite{Ber86}. Moreover, the screening
current $S_n(z)$ is fermionic and the others are bosonic.


\section{Discussions}
\label{Con} \setcounter{equation}{0}

We have constructed the explicit expressions of the free-field
representation for the $osp(2n|2n)$ current algebra at an
arbitrary level $k$, and the corresponding energy-momentum tensor.
We have also found the free-field representation of the $2n$
associated screening currents of the first kind.

The free-field realization (\ref{Free-D-1})-(\ref{Free-D-2}) of
the $osp(2n|2n)$ current algebra gives rise to the Fock
representation of the corresponding current algebra in terms of
the free fields (\ref{OPE-F-1})-(\ref{OPE-F-2}). It provides
explicit realizations of the  vertex operator construction
\cite{Lep84,Pri02} of representations for affine superalgebra
$osp(2n|2n)_{k}$. Moreover, these representations are in general
not irreducible for the current algebra. To obtain irreducible
representations, one needs the associated screening charges, which
are the integrals of the corresponding screening currents
(\ref{Screen-D-1})-(\ref{Screen-D-2}) and performs the cohomology
analysis as in \cite{Fat86,Fei90,Bou90,Ber89}.

To fully take the advantage of the CFT method, one needs to
construct its primary fields. It is well-known that there exist
two types of representations for the underlying finite-dimensional
superalgebra $osp(2n|2n)$: typical and atypical representations.
Atypical representations have no counterpart in the bosonic
algebra setting and the understanding of such representations is
still very much incomplete. Although the construction of the
primary fields associated with typical representations is similar
to the bosonic algebra cases, it is a highly nontrivial task to
construct the primary fields associated with atypical
representations \cite{Zha05}.

\section*{Acknowledgements}
The financial support from  the Australian Research Council is
gratefully acknowledged. WLY has also been partially supported by
the New Staff Research Grant of the University of Queensland.

\section*{Appendix A: Defining representation of $osp(2n|2n)$
} \setcounter{equation}{0}
\renewcommand{\theequation}{A.\arabic{equation}}

Let $V$ be a $\Zb_2$-grading $4n$-dimensional vector space with an
orthonornal basis $\{|i\rangle, i=1,\ldots, 4n\}$. The
$\Zb_2$-grading is chosen as: $[1]=\cdots=[2n]=0,\,
[2n+1]=\cdots=[4n]=1.$ Let  $e_{ij}$, $i,j=1,\ldots,n$, be an
$n\times n$ matrix with entry $1$ at the $i$th row and the $j$th
column and zero elsewhere. The $4n$-dimensional defining
representation of $osp(2n|2n)$, denoted by $\rho_{0}$, is given by
the following $4n\times 4n$ matrices, \bea \hspace{-1.26truecm}
 \rho_{0}\lt(E_{\d_m-\d_l}\rt)
   \hspace{-0.32truecm}&=&\hspace{-0.32truecm}
   \left(\begin{array}{c|c}{\,\,\,\,\,\,}&{\,\,\,\,\,\,}\\[6pt]
   \hline &{\begin{array}{cc}e_{ml}&\\
   &-e_{lm}\end{array}}\end{array}\right),\,\,
 \rho_{0}\lt(F_{\d_m-\d_l}\rt)=
   \left(\begin{array}{c|c}{\,\,\,\,\,\,}&{\,\,\,\,\,\,}\\[6pt]
   \hline &{\begin{array}{cc}e_{lm}&\\
   &-e_{ml}\end{array}}\end{array}\right),\,\, m<l,\label{F-R-1}\\[8pt]
\hspace{-1.26truecm}
 \rho_{0}\lt(E_{2\d_l}\rt)
 \hspace{-0.32truecm}&=&\hspace{-0.32truecm}
   \left(\begin{array}{c|c}{\,\,\,\,\,\,}&{\,\,\,\,\,\,}\\[6pt]
   \hline &{\begin{array}{cc}0&e_{ll}\\
   0&0\end{array}}\end{array}\right),\,\,
 \rho_{0}\lt(F_{2\d_l}\rt)=
   \left(\begin{array}{c|c}{\,\,\,\,\,\,}&{\,\,\,\,\,\,}\\[6pt]
   \hline &{\begin{array}{cc}0&0\\
   e_{ll}&0\end{array}}\end{array}\right),\\[8pt]
\hspace{-1.26truecm}
 \rho_{0}\lt(E_{\d_m+\d_l}\rt)
   \hspace{-0.32truecm}&=&\hspace{-0.32truecm}
   \left(\begin{array}{c|c}{\,\,\,\,\,\,}&{\,\,\,\,\,\,}\\[6pt]
   \hline &{\begin{array}{cc}0&e_{ml}+e_{lm}\\
   0&0\end{array}}\end{array}\right),\,\,
 \rho_{0}\lt(F_{\d_m+\d_l}\rt)=
   \left(\begin{array}{c|c}{\,\,\,\,\,\,}&{\,\,\,\,\,\,}\\[6pt]
   \hline &{\begin{array}{cc}0&0\\
   e_{ml}+e_{lm}&0\end{array}}\end{array}\right),\,\, m<l,\\[8pt]
\hspace{-1.26truecm}
 \rho_{0}\lt(E_{\d_l-\e_i}\rt)
   \hspace{-0.32truecm}&=&\hspace{-0.32truecm}
   \left(\begin{array}{c|c}&{\begin{array}{cc}0&0\\
   0&e_{il}\end{array}}\\[6pt]
   \hline {\begin{array}{cc}e_{li}&0\\0&0\end{array}}&\\
   \end{array}\right),\,\,
 \rho_{0}\lt(F_{\d_l-\e_i}\rt)=
   \left(\begin{array}{c|c}&{\begin{array}{cc}e_{il}&0\\
   0&0\end{array}}\\[6pt]
   \hline {\begin{array}{cc}0&0\\0&-e_{li}\end{array}}&\\
   \end{array}\right),\\[8pt]
\hspace{-1.26truecm}
 \rho_{0}\lt(E_{\d_l+\e_i}\rt)
   \hspace{-0.32truecm}&=&\hspace{-0.32truecm}
   \left(\begin{array}{c|c}&{\begin{array}{cc}0&e_{il}\\
   0&0\end{array}}\\[6pt]
   \hline {\begin{array}{cc}0&e_{li}\\0&0\end{array}}&\\
   \end{array}\right),\,\,
 \rho_{0}\lt(F_{\d_l+\e_i}\rt)=
   \left(\begin{array}{c|c}&{\begin{array}{cc}0&0\\
   e_{il}&0\end{array}}\\[6pt]
   \hline {\begin{array}{cc}0&0\\-e_{li}&0\end{array}}&\\
   \end{array}\right),\\[8pt]
\hspace{-1.26truecm}
 \rho_{0}\lt(E_{\e_i-\e_j}\rt)
   \hspace{-0.32truecm}&=&\hspace{-0.32truecm}
   \left(\begin{array}{c|c}{\begin{array}{cc}e_{ij}&\\&-e_{ji}\end{array}}&\\
   [8pt]\hline {\,\,\,\,\,\,}&{\,\,\,\,\,\,}\end{array}\right),\,\,
 \rho_{0}\lt(F_{\e_i-\e_j}\rt)=
   \left(\begin{array}{c|c}{\begin{array}{cc}e_{ji}&\\&-e_{ij}\end{array}}&\\
   [8pt]\hline {\,\,\,\,\,\,}&{\,\,\,\,\,\,}\end{array}\right),\qquad\quad i<j,\\[8pt]
\hspace{-1.26truecm}
 \rho_{0}\lt(E_{\e_i+\e_j}\rt)
   \hspace{-0.32truecm}&=&\hspace{-0.32truecm}
   \left(\begin{array}{c|c}{\begin{array}{cc}0&e_{ij}-e_{ji}\\0&0\end{array}}&\\
   [8pt]\hline {\,\,\,\,\,\,}&{\,\,\,\,\,\,}\end{array}\right),\,\,
 \rho_{0}\lt(F_{\e_i+\e_j}\rt)=
   \left(\begin{array}{c|c}{\begin{array}{cc}0&0\\-e_{ij}+e_{ji}&0\end{array}}&\\
   [8pt]\hline {\,\,\,\,\,\,}&{\,\,\,\,\,\,}\end{array}\right),\,\, i<j,\\[8pt]
\hspace{-1.26truecm}
 \rho_{0}\lt(H_{\d_m-\d_l}\rt)
   \hspace{-0.32truecm}&=&\hspace{-0.32truecm}
   \left(\begin{array}{c|c}{\,\,\,\,\,\,}&{\,\,\,\,\,\,}\\[6pt]
   \hline &{\begin{array}{cc}e_{mm}-e_{ll}&\\
   &e_{ll}-e_{mm}\end{array}}\end{array}\right),\qquad\qquad m<l,\\[8pt]
\hspace{-1.26truecm}
 \rho_{0}\lt(H_{\d_m+\d_l}\rt)
   \hspace{-0.32truecm}&=&\hspace{-0.32truecm}
   \left(\begin{array}{c|c}{\,\,\,\,\,\,}&{\,\,\,\,\,\,}\\[6pt]
   \hline &{\begin{array}{cc}e_{mm}+e_{ll}&\\
   &-e_{mm}-e_{ll}\end{array}}\end{array}\right),\qquad\qquad m<l,\\[8pt]
\hspace{-1.26truecm}
 \rho_{0}\lt(H_{2\d_l}\rt)
   \hspace{-0.32truecm}&=&\hspace{-0.32truecm}
   \left(\begin{array}{c|c}{\,\,\,\,\,\,}&{\,\,\,\,\,\,}\\[6pt]
   \hline &{\begin{array}{cc}e_{ll}&\\
   &-e_{ll}\end{array}}\end{array}\right),\\[8pt]
\hspace{-1.26truecm}
 \rho_{0}\lt(H_{\d_l-\e_i}\rt)
   \hspace{-0.32truecm}&=&\hspace{-0.32truecm}
   \left(\begin{array}{c|c}{\begin{array}{cc}e_{ii}&\\&-e_{ii}\end{array}}&\\[6pt]
   \hline &{\begin{array}{cc}e_{ll}&\\
   &-e_{ll}\end{array}}\end{array}\right),\,\,
 \rho_{0}\lt(H_{\d_l+\e_i}\rt)\hspace{-0.1truecm}=\hspace{-0.1truecm}
   \left(\begin{array}{c|c}{\begin{array}{cc}-e_{ii}&\\&e_{ii}\end{array}}&\\[6pt]
   \hline &{\begin{array}{cc}e_{ll}&\\
   &-e_{ll}\end{array}}\end{array}\right),\\[8pt]
\hspace{-1.26truecm}
 \rho_{0}\lt(H_{\e_i-\e_j}\rt)
   \hspace{-0.32truecm}&=&\hspace{-0.32truecm}
   \left(\begin{array}{c|c}{\begin{array}{cc}e_{ii}-e_{jj}&\\&e_{jj}-e_{ii}\end{array}}&\\[6pt]
   \hline {\,\,\,\,}&{\,\,\,\,}\end{array}\right),\qquad\qquad i<j,\\[8pt]
\hspace{-1.26truecm}
 \rho_{0}\lt(H_{\e_i+\e_j}\rt)
   \hspace{-0.32truecm}&=&\hspace{-0.32truecm}
   \left(\begin{array}{c|c}{\begin{array}{cc}e_{ii}+e_{jj}&\\&-e_{ii}-e_{jj}\end{array}}&\\[6pt]
   \hline {\,\,\,\,}&{\,\,\,\,}\end{array}\right),\qquad\qquad
   i<j. \label{F-R-2}
\eea We introduce $2n$ linear-independent generators $H_i$
$(i=1,\ldots 2n)$, \bea
 H_l&=&H_{2\d_l},\qquad 1\leq l\leq n,\label{D-H-1}\\
 H_{n+i}&=&\frac{1}{2}(H_{\e_i-\e_j}+H_{\e_i+\e_j}),\qquad i=1,\ldots,
   n-1,\,{\rm and}\,\,i<j,\\
 H_{2n}&=&\frac{1}{2}\lt(H_{\e_i+\e_{n}}-H_{\e_i-\e_{n}}\rt), \qquad i\leq n-1.\label{D-H-2}
\eea Actually, the above generators $\{H_i\}$ span the Cartan
subalgebra of $osp(2n|2n)$. In the defining representation, these
generators can be realized by \bea
 \rho_{0}\lt(H_l\rt)&=&
   \left(\begin{array}{c|c}{\,\,\,\,\,\,}&{\,\,\,\,\,\,}\\[6pt]
   \hline &{\begin{array}{cc}e_{ll}&\\
   &-e_{ll}\end{array}}\end{array}\right),\,\,
   l=1,\ldots,n,\\[8pt]
 \rho_{0}\lt(H_{n+i}\rt)&=&
   \left(\begin{array}{c|c}{\begin{array}{cc}e_{ii}&\\
   &-e_{ii}\end{array}}\\[6pt]
   \hline {\,\,\,\,} &{\,\,\,\,}\end{array}\right),\,\,
   i=1,\ldots,n.
\eea

The corresponding nondegenerate invariant bilinear supersymmetric
form of $osp(2n|2n)$ is given by \bea
 (x,y)=\frac{1}{2}str\lt(\rho_0(x)\rho_0(y)\rt),\qquad
  \forall x,y\in osp(2n|2n).\label{Bilinear-D}
\eea Here  the supertrace, for any $4n\times 4n$ matrix $A$, is
defined by \bea
 str\lt(A\rt)=\sum_{l=1}^{4n}(-1)^{[l]}A_{ll}
 =\sum_{l=1}^{2n}A_{ll}-\sum_{l=2n+1}^{4n}A_{ll}.
\eea


\end{document}